\documentclass[11pt]{article}

\usepackage[T1]{fontenc}
\usepackage[utf8]{inputenc}
\usepackage{lmodern}   
\usepackage[margin=1in]{geometry}
\usepackage{microtype}
\usepackage{graphicx}
\usepackage{booktabs}
\usepackage{array}
\usepackage{amsmath}
\usepackage{enumitem}
\usepackage{xcolor}
\usepackage[hidelinks]{hyperref}
\usepackage{caption}

\newcommand{\code}[1]{\texttt{#1}}

\title{\textbf{Zengram-Lite: An In-Browser Agentic-Memory Framework}\\[0.3em]
\large Semantic Knowledge, Session Tracking, and Token-Budgeted Context}
\author{Gene Zhang\\[0.2em]
\normalsize\href{mailto:genegzhang@gmail.com}{genegzhang@gmail.com}\\[0.2em]
\normalsize\url{https://github.com/genezhang/zengram-lite}}
\date{2026}

\begin{document}
\maketitle

\begin{abstract}
AI agents increasingly run in the browser, and they need somewhere to keep what
they learn. The client-side state of the art, however, is a vector index ---
nearest-neighbor search over embeddings --- with the rest of an agent's memory
left to application code: the conversation history is an array in
\code{localStorage}, context management is hand-rolled truncation, and there is
no shared notion of a fact's confidence, its provenance, or the session that
produced it. A vector store is not a memory \emph{framework}.

We present \textbf{zengram-lite}, an agentic-memory framework compiled to a
single $\sim$2.95\,MB gzipped WebAssembly artifact that runs entirely in a
browser tab. It provides three tiers over one transactional store. The
\textbf{knowledge} tier offers hybrid vector-and-full-text recall with a fact
\emph{lifecycle} --- confidence that rises and falls as facts are confirmed or
contradicted, importance that decays over time, supersession that treats a
restated fact as an update rather than a duplicate, and scope namespacing. The
\textbf{session-tracking} tier models the conversation itself as first-class
data: sessions, turns, typed content parts, and tool calls with state and
timing, all as queryable tables rather than an opaque message log. The
\textbf{context-assembly} tier turns that structure into a token-budgeted prompt
through a six-phase assembly that packs system instructions, knowledge, and
recent history under a caller's budget, and returns a stable fingerprint that
signals when a prompt prefix can be reused. The bundle is a \textbf{superset} of
the zeta-lite SQL engine~[ZL] --- the same \code{.wasm} re-exports the full
Postgres-compatible surface, MVCC snapshot isolation, and copy-on-write database
branching --- so memory inherits transactional consistency across all three
tiers and branchable state for free. Because the wasm surface is synchronous
while the framework's canonical operations depend on an asynchronous LLM and
embedder, zengram-lite exposes a \textbf{bring-your-own-result} seam that runs
the framework's real code paths over results the application computes in
JavaScript. This is a v0.1 preview; \S8 states its boundaries plainly.
\end{abstract}

\begin{center}
\parbox{0.92\linewidth}{\small
\textit{Database management system engines}; \textit{Information retrieval query
processing}; $\bullet$\,\textit{Computing methodologies} $\rightarrow$
\textit{Intelligent agents}; \textit{Knowledge representation and reasoning}.

}
\end{center}

\section*{Availability}
\addcontentsline{toc}{section}{Availability}

The compiled zengram-lite bundle is distributed as a GitHub Release with the
web-target bundle attached (publication to npm as \code{zengram-lite} is pending);
it is free for any use, including commercial. The Zengram framework source is
planned for open-source release under Apache-2.0 (publication pending); the Zeta
engine source stays closed, so the WebAssembly build --- which links both --- is
\emph{distributed} as a prebuilt binary rather than itself open source. The
hand-authored surface --- three
interactive demo pages (an agent-memory demo, the full SQL playground, and a
local AI agent that runs its loop, tools, and memory in the tab), the API
reference, and the smoke/e2e harnesses used in this paper --- is public at
\url{https://github.com/genezhang/zengram-lite}. Every measurement in \S7 is
reproducible from that repository against the \emph{published} artifact: the
artifact size (\code{gzip -c demo/pkg-web/zengram\_wasm\_bg.wasm | wc -c}), the
memory-core and agent-loop smoke tests (\code{node demo/smoke.mjs}, \code{node
demo/smoke\_agent.mjs}), the real-model recall test (\code{node
demo/smoke\_model.mjs}), and the headless-browser DOM test (\code{node
demo/e2e\_ops.mjs}). The companion engine report is~[ZL].

\section{Introduction}

Agents have moved into the browser. A growing class of applications run an agent
loop --- plan, call tools, observe, repeat --- inside a tab, for the same reasons
data moved client-side generally: privacy, offline operation, and the absence of
a server round-trip between a thought and the state that records it. Such an
agent needs memory. It needs to remember facts it has learned and recall them
later by meaning; it needs to keep track of the conversation it is having, the
tools it has called and how they turned out; and it needs, on every turn, to
assemble a prompt that fits a model's context window out of far more material
than will fit.

The dominant answer today is a vector store. Libraries embed text with a model
running in the tab (transformers.js, ORT-Web) and index the vectors for
nearest-neighbor search, typically over IndexedDB. This is genuinely useful and
it is the part of agent memory that is well served in the browser. But it is only
one part. A vector store has no notion of a fact's confidence or whether it has
since been contradicted; it does not model the conversation, so the session
history, the turns, and the tool calls are left to the application to keep in
ad-hoc structures; and it offers nothing for the actual hard problem of each turn
--- deciding what to put in the prompt and what to leave out under a token
budget. In practice a browser agent assembles its memory from a vector index plus
a pile of hand-written glue, and the glue is where the framework should be.

Server-hosted agent-memory frameworks do model these things. Systems in the mem0
/ MemGPT / Letta lineage provide a fact store, conversational memory, and context
management as a coherent library --- but they run in a Python or Node process on
a server, which puts both the framework and the user's data back on the far side
of a network. For a browser agent that exists precisely to keep computation and
data on the device, that is the wrong deployment.

We take a different bet. \textbf{Zengram-lite} compiles a complete agentic-memory
framework to WebAssembly and runs it entirely in the tab, over a transactional
SQL engine that ships in the same artifact. It is the browser form factor of the
full Zengram framework, and it is a \textbf{superset} of the zeta-lite database
engine~[ZL]: the glue crate \code{zengram-wasm} links the engine and the
framework into one cdylib, so a single \code{.wasm} exports both the full SQL
surface (\code{ZetaDb}/\code{ZetaTxn}/\code{ZetaCursor}) and the memory tier
(\code{ZengramMemory}). A page that needs memory and SQL loads one engine, not
two. On top of that engine the framework provides three tiers --- a knowledge
tier with a fact lifecycle, a session-tracking tier that models the conversation
as data, and a context-assembly tier that produces a token-budgeted prompt ---
each backed by real tables the framework keeps consistent.

\textbf{Contributions.}
\begin{enumerate}[leftmargin=1.4em,itemsep=0.2em]
  \item \textbf{A knowledge tier that is a system, not a store} (\S4): hybrid
    vector-and-full-text recall with a fact lifecycle --- confidence adjusted by
    confirm/contradict, importance that decays with a half-life, supersession of
    restated facts, scope namespacing, and provenance back to the turn that
    produced a fact.
  \item \textbf{First-class, client-side session tracking and token-budgeted
    context assembly} (\S5): sessions, turns, typed parts, and tool calls as
    queryable tables, and a six-phase \code{assembleContext} that packs a prompt
    under a token budget and returns a stable reuse fingerprint. This is the tier
    that lets a browser agent run its whole loop in the tab, and the one no other
    in-browser memory system provides.
  \item \textbf{Running an LLM-dependent framework on a synchronous wasm surface}
    (\S6): the \textbf{bring-your-own-result} seam that resolves the mismatch
    between a synchronous wasm API and the asynchronous LLM/embedder the
    framework's canonical operations need, by running the framework's real code
    paths over results the application computes in JavaScript.
  \item \textbf{The superset embodiment} (\S3): the whole framework plus a full
    transactional SQL engine with copy-on-write branching in one $\sim$2.95\,MB
    artifact, with a choice of isolated or shared engine, over the zeta-lite
    core~[ZL].
  \item \textbf{Position in the Zeta/Zengram family}: the smallest, most complete
    public form of the full framework, and the in-browser teaser for it.
\end{enumerate}

Throughout, the argument is that the missing piece on the client side was never
the vector index --- that part is solved --- but the \emph{framework} around it:
the lifecycle, the session model, and the context assembly that together make
stored vectors behave like memory. Zengram-lite is a v0.1 preview; \S8 is
explicit about what is and is not wired yet.

\section{Background and Related Work}

\textbf{Agent-memory frameworks.} A line of systems treats an agent's memory as a
first-class component rather than a bare vector index. MemGPT / Letta~[1] frame
the context window as a managed resource, paging information between an in-context
working set and external storage; mem0~[2] provides a fact-extraction-and-recall
memory layer for LLM applications; retrieval frameworks such as LangChain and
LlamaIndex~[3] bundle conversational memory and context construction. These share
the shape zengram-lite adopts --- a store plus a lifecycle plus context
management --- but they are built to run server-side, in a Python or Node
process, with the store frequently a hosted vector database. Zengram-lite's
difference is deployment: the \emph{entire} framework, store included, runs in the
browser tab, so an agent's memory and the user's data never leave the device.

\textbf{Vector search in the browser.} The client-side building block that is
well established is embedding-plus-nearest-neighbor. Embedding models run in the
tab via transformers.js or ONNX Runtime Web, and vectors are indexed with in-wasm
HNSW libraries or stored in IndexedDB. These are the right tool for semantic
recall, and zengram-lite uses exactly this class of embedder at its boundary
(\S6.2). What they are not is a memory framework: they carry no fact lifecycle, no
session or turn model, and no context assembly. They are the substrate the
knowledge tier sits on, not a substitute for the framework.

\textbf{Client-side agent state as it is built today.} Absent a framework, a
browser agent keeps its conversation in application structures --- an array of
messages in \code{localStorage} or IndexedDB --- and manages context by hand,
typically truncating to the last $k$ turns. Tool calls are logged, if at all, in
the same ad-hoc way, with no shared schema tying a stored fact to the turn and
tool call that produced it. This is the status quo zengram-lite is meant to
replace with a coherent, queryable model.

\textbf{The engine.} Zengram-lite is a superset of \textbf{zeta-lite}~[ZL], whose
report is the account of the underlying database: a log-centric asynchronous MVCC
engine that sustains overlapping snapshot-isolated transactions on a single wasm
thread, exposes a feature-complete Postgres surface (JSONB with GIN, full-text
search, HNSW vector search, SQL/PGQ graph queries, multi-database), offers
copy-on-write whole-database branching, and persists via snapshot to the Origin
Private File System without a worker, \code{SharedArrayBuffer}, or
cross-origin-isolation headers. We summarize what the framework inherits in \S3
and refer to~[ZL] for the mechanism and its evaluation; this paper does not
re-argue the engine.

\textbf{The delta.} To our knowledge, no in-browser system offers a coherent
agentic-memory framework --- a knowledge tier with a lifecycle, first-class
session tracking, and token-budgeted context assembly --- as a single client-side
artifact. Zengram-lite provides one, over a transactional engine, in
$\sim$2.95\,MB.

\section{The Zeta Substrate}

\begin{figure}[htbp]
  \centering
  \includegraphics[width=\linewidth]{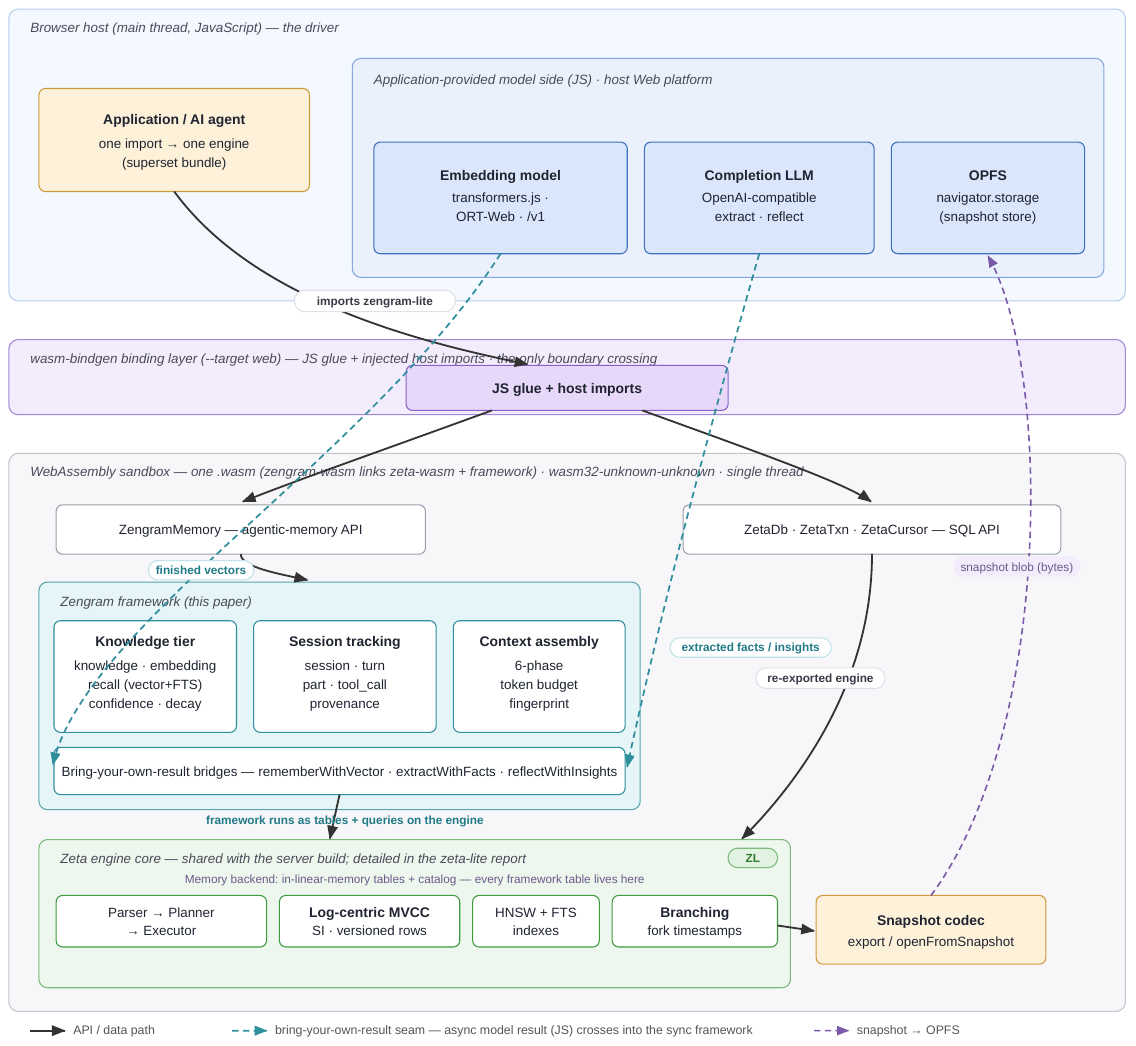}
  \caption{Zengram-lite superset architecture: the Zengram framework tiers (teal)
  as tables and queries over the shared Zeta engine core (green, detailed in the
  zeta-lite report), inside one wasm module that also re-exports the SQL surface.
  The bring-your-own-result seam (teal, dashed) carries async model results from
  the JS side into the synchronous framework.}
  \label{fig:arch}
\end{figure}

Zengram-lite is a framework, but it is not a framework in isolation: it runs over
the Zeta database engine, and much of what makes it viable in a browser is
inherited from that engine rather than built anew. This section states what is
inherited and points to the companion report~[ZL] for the account of how; it
deliberately does not re-derive the engine. Figure~\ref{fig:arch} gives the whole
picture --- the two API surfaces the one \code{.wasm} exports, the framework
tiers, the shared engine core, and the seams that cross into the browser host.

\subsection{What zengram-lite inherits}

From zeta-lite~[ZL], the framework gets, without adding code of its own:
\begin{itemize}[leftmargin=1.4em,itemsep=0.2em]
  \item \textbf{A transactional store with snapshot isolation.} The memory tiers
    are tables in a log-centric MVCC engine, so a write that spans knowledge,
    session, and provenance tables is one consistent transaction rather than a
    multi-store reconciliation problem.
  \item \textbf{The retrieval primitives.} HNSW vector indexing and full-text
    search are engine features; the knowledge tier's hybrid recall (\S4.2) is a
    query over them, not a separate index the framework maintains.
  \item \textbf{Copy-on-write branching.} The engine can fork the whole database
    at the cost of a timestamp~[ZL~\S3.6]. Memory can therefore be branched ---
    an agent can explore a speculative line of memory and merge or discard it ---
    which the framework surfaces as in-bundle memory branching (exposed in a later
    release; \S8).
  \item \textbf{Snapshot-to-OPFS durability.} The whole database serializes to a
    byte blob and rehydrates from one, on the main thread, with no worker or
    special headers~[ZL~\S5]. Memory persistence (\S6.4) is this mechanism.
  \item \textbf{A small, headers-free artifact.} The engine is $\sim$2.87\,MB
    gzipped and runs in any browser context with OPFS. The framework adds little
    on top (\S7.3).
\end{itemize}

When this paper says ``the store,'' ``recall,'' or ``branching,'' it means the
engine's, via~[ZL]. The novelty argued here is the framework above them.

\subsection{The superset bundle}

The glue crate \code{zengram-wasm} links \code{zeta-wasm} and the Zengram
framework into a single cdylib, so one compiled \code{.wasm} exports the full SQL
engine (\code{ZetaDb}/\code{ZetaTxn}/\code{ZetaCursor}) \emph{and} the memory tier
(\code{ZengramMemory}). This is the concrete meaning of ``superset'': a page that
needs both a SQL database and agent memory imports one package and instantiates
one engine, with no double-load and no second wasm heap. zeta-lite remains the
lean, SQL-only package for pages that need only the database; zengram-lite is the
strict superset for pages that also need memory. The cost of the framework tier on
top of the engine is small --- \S7.3 measures it at roughly 0.08\,MB gzip over the
engine's 2.87\,MB.

\subsection{Why memory over a SQL engine}

Building agent memory over a transactional SQL engine, rather than as a bespoke
store, is what makes the three tiers a single system. The knowledge tier is HNSW
and full-text indexes over the \code{knowledge} and \code{embedding} tables; the
session tier is rows in \code{session}, \code{turn}, \code{part}, and
\code{tool\_call}; provenance is foreign-key-shaped relationships among them. A
recall, a turn append, and the fact-extraction that links new knowledge back to
the turn it came from all commit in the same engine, under the same snapshot
isolation. The alternative --- a vector database beside a document store beside a
hand-rolled session log --- has no single point of consistency; here there is one.
This is the payoff of putting the framework on the engine rather than beside it.

\subsection{The schema as one system}
\label{sec:schema}

\begin{figure}[htbp]
  \centering
  \includegraphics[width=\linewidth]{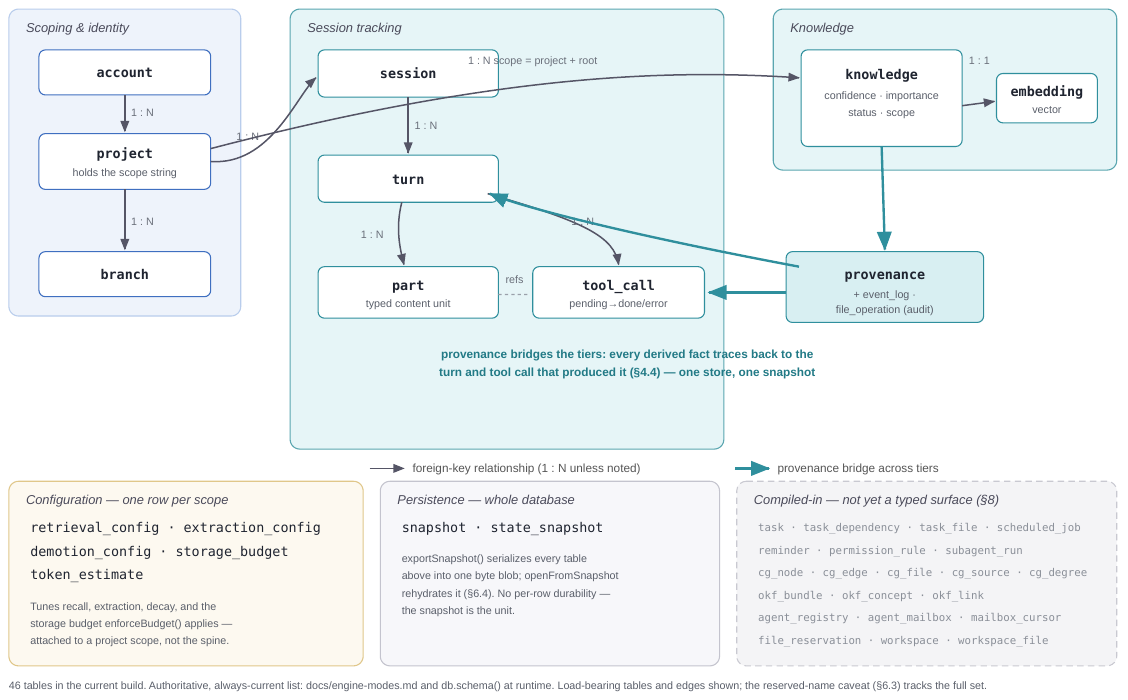}
  \caption{The framework schema, by tier. A foreign-key spine runs left to right
  --- an account owns projects (which carry the scope string), a project owns
  sessions, and a session owns turns, each turn owning its typed parts and tool
  calls. Knowledge is scoped to a project (plus the root scope). The teal
  provenance bridge is the point: \code{provenance} ties a derived knowledge fact
  back to the exact turn and tool call that produced it, across tier boundaries,
  in one store under one snapshot. Configuration, persistence, and the
  compiled-in-but-not-yet-exposed tables sit apart from the spine.}
  \label{fig:schema}
\end{figure}

The concrete form of that payoff is the framework's schema, and it is worth seeing
as a whole because it is the evidence for the claim that zengram-lite is a
coherent system rather than a pile of tables. The current build defines 46 tables;
Figure~\ref{fig:schema} shows the load-bearing ones and the relationships among
them, grouped by the tiers of \S4--\S5.

The spine is ordinary relational structure: \code{account} $\to$ \code{project}
$\to$ \code{session} $\to$ \code{turn} $\to$ \code{part} / \code{tool\_call}, with
\code{knowledge} scoped to a project and its \code{embedding} alongside. What makes
it a memory framework rather than a set of logs is the \textbf{provenance bridge}
--- the teal edges in the figure. When the extraction seam (\S6.2) stores a fact
derived from a turn, \code{provenance} records which turn and which tool call it
came from, so a recalled fact carries its origin across the tier boundary between
knowledge and session. Because every one of these tables lives in the same MVCC
store, that link is a foreign key under one snapshot, not a join across separate
systems that can disagree.

Three groups sit off the spine. \textbf{Configuration} tables
(\code{retrieval\_config}, \code{extraction\_config}, \code{demotion\_config},
\code{storage\_budget}, \code{token\_estimate}) hold one row per scope and tune
recall, extraction, decay, and the storage budget that \code{enforceBudget}
applies (\S5.3). \textbf{Persistence} (\code{snapshot}, \code{state\_snapshot}) is
the whole-database snapshot codec of \S6.4 --- \code{exportSnapshot} serializes
every table above into one blob. And a substantial set of tables is
\textbf{compiled in but not yet exposed} as a typed surface (\S8): task
management, reminders, permission rules, subagent runs, a code graph, an
object-knowledge format, and fleet/mailbox coordination. That the build already
carries these --- reachable via SQL today, typed later --- indicates the scope of
the full Zengram framework this teaser fronts. The authoritative,
always-current table list is in the repository (\code{docs/engine-modes.md}) and
reachable at runtime via \code{db.schema()}; the set grows with the framework, and
the reserved-name caveat of \S6.3 tracks it.

\section{The Knowledge Tier}

The knowledge tier is what most systems mean by ``agent memory'': store facts,
recall them by meaning. Zengram-lite's difference is that a fact is not a frozen
vector but a row with a lifecycle.

\subsection{What a fact is}

A fact is a row in the \code{knowledge} table. It carries a \textbf{subject} (a
short label such as \code{"editor prefs"}), the \textbf{content} (the fact as a
sentence), a \textbf{scope} (a free-form namespace string such as
\code{"agent/session-42"}), a \textbf{confidence} and an \textbf{importance} (each
in $0..1$), a lifecycle \textbf{status}, counters for how many times it has been
confirmed and contradicted, and update timestamps that drive decay. Recall is
always within a scope, so scopes partition an agent's memory into namespaces ---
per project, per session, per user --- over one store.

\subsection{Remember and recall}

\code{remember(subject, content, scope)} embeds the fact --- as the string
\code{"subject: content"} --- and stores it, returning a knowledge id.
\code{recall(query, scope)} runs a \textbf{hybrid vector-and-full-text} search
over the scope's facts and returns them ranked by meaning, each hit carrying a
similarity score and the fact's importance. The vector half is the point: a query
with no keyword overlap with a stored fact still retrieves it. \S7.1 shows this
with real all-MiniLM vectors --- asking ``what hot beverage do I like?'' recalls a
stored ``drinks green tea'' fact that shares no words with the query.

Facts with the \textbf{same subject in a scope are treated as updates}, not
duplicates: restating a subject supersedes the prior fact rather than accumulating
a near-copy. The application stores distinct subjects for distinct facts and lets
supersession handle restatement. \code{remember} stores content \textbf{verbatim}
--- it does not split prose into atomic facts on its own; automatic extraction is
the bring-your-own-result path of \S6.2.

\subsection{The lifecycle: confidence, importance, decay}

What makes the tier \emph{memory} rather than a static index is that facts change
standing over time, through deterministic operations that need no model and run
before an embedder is even wired:
\begin{itemize}[leftmargin=1.4em,itemsep=0.2em]
  \item \textbf{\code{confirm(id)}} --- the fact was seen or validated again; its
    confidence rises.
  \item \textbf{\code{contradict(id)}} --- the fact turned out wrong; its
    confidence falls.
  \item \textbf{\code{decay(halfLifeDays)}} --- importance decays exponentially
    with the given half-life, charging only the time since each fact's last decay,
    and returns the number of facts touched. Run periodically, it fades stale
    memory so the store does not grow without bound over months.
\end{itemize}

These are the operations that distinguish a memory from a log: a fact the agent
keeps re-encountering strengthens, one the world contradicts weakens, and one
nothing touches quietly fades. \code{factsAboutPeer(peer, limit)} returns the
active facts attributed to a given peer, most important first, for memory that is
about \emph{who} said or did something, not only \emph{what}.

\subsection{Provenance}

A fact derived from a conversation carries where it came from. When the
fact-extraction bridge (\S6.2) stores facts from a turn, it records session and
turn provenance on each, so a recalled fact can be traced back to the exact turn
that produced it. The framework compiles a fuller provenance-tracing surface
(\code{traceBack}/\code{traceForward}) that a later release will expose typed
(\S8); in v0.1 the provenance is recorded and inspectable via SQL.

\section{The Session-Tracking Tier}

This is the tier that separates a memory framework from a vector store, and the
one that makes ``run the agent in the tab'' tractable. It models the conversation
itself as data, and it turns that data into a prompt.

\subsection{The conversation as data}

An agent's transcript is usually an opaque array of messages. Zengram-lite models
it as a small relational structure the framework keeps consistent:
\begin{itemize}[leftmargin=1.4em,itemsep=0.2em]
  \item A \textbf{session} (\code{session}) is one run of the agent, created under
    a project scope, with its own trunk branch.
  \item A \textbf{turn} (\code{turn}) is one exchange --- user or assistant ---
    carrying its role, lifecycle status, input/output token counts, cost, and
    finish reason.
  \item A \textbf{part} (\code{part}) is an atomic, typed content unit of a turn
    --- a text part, a tool part --- with a JSON payload and a position, so a
    turn's content is structured rather than a single string.
  \item A \textbf{tool call} (\code{tool\_call}) records a tool invocation with a
    state machine (pending $\to$ running $\to$ completed or error), a category,
    target paths, timing, and token cost.
\end{itemize}

Because these are real tables, the transcript is queryable: ``which tool calls
errored in this session,'' ``how many tokens did this branch cost,'' ``what parts
made up that turn'' are SQL, not application bookkeeping.

\subsection{The loop, in the bundle}

Together these let a browser agent run its entire loop through the framework:
\code{createSession} opens a session and its branch; \code{appendTurn} and
\code{addPart} record each exchange; \code{recordToolCall} and
\code{completeToolCall} bracket every tool use; \code{assembleContext} (\S5.3)
builds the next prompt; and \code{extractWithFacts} / \code{reflectWithInsights}
(\S6.2) fold what was learned back into the knowledge tier. The
\code{smoke\_agent.mjs} harness walks exactly this sequence against the real
artifact and checks it end to end (\S7.1). None of it requires a server: the
record of the conversation, the prompt assembly, and the derived memory all live
in the tab.

\subsection{Token-budgeted context assembly}
\label{sec:assemble}

\begin{figure}[htbp]
  \centering
  \includegraphics[width=\linewidth]{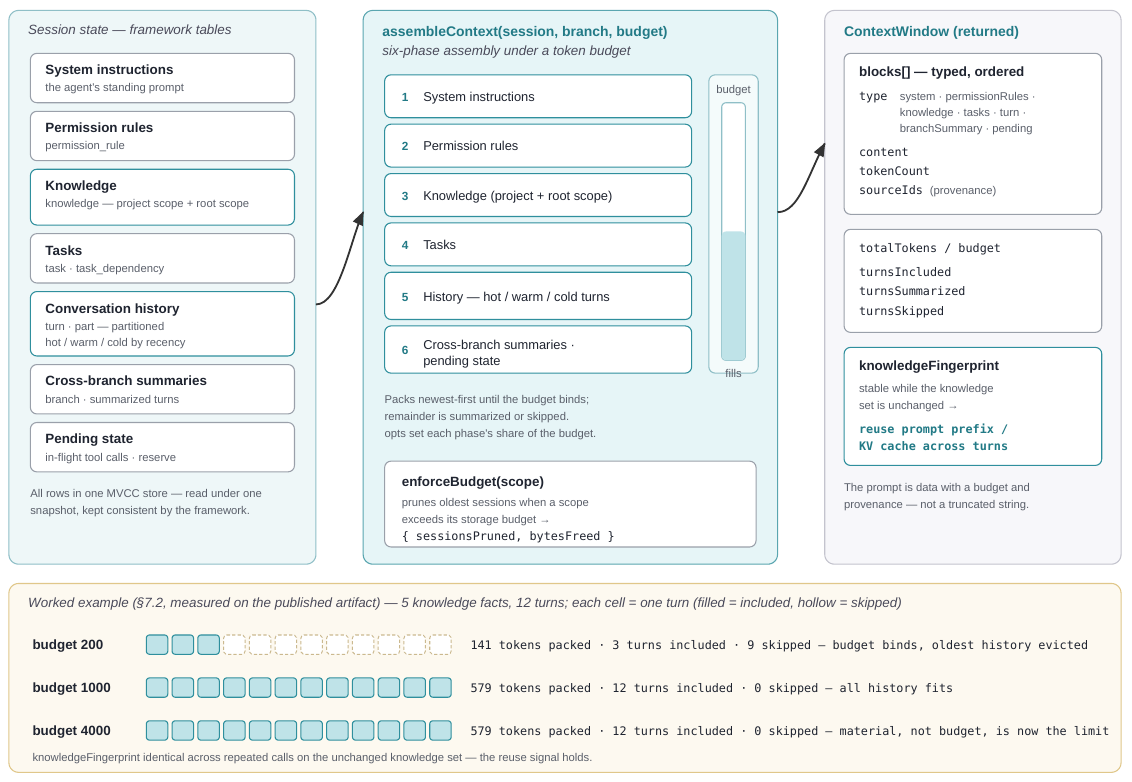}
  \caption{Token-budgeted context assembly. \code{assembleContext} reads the
  session-state tables under one snapshot, packs them through six phases
  newest-first until the token budget binds, and returns a typed
  \code{ContextWindow} with per-block provenance and a reuse fingerprint. The
  bottom strip is the \S7.2 measurement: at budget 200 only 3 of 12 turns are
  included (9 evicted); at budget 1000 and 4000 all 12 fit.}
  \label{fig:assemble}
\end{figure}

The operation an agent performs most and a vector store helps with least is
building the next prompt. On every turn the agent has more material --- system
instructions, relevant knowledge, recent history, pending tool state --- than fits
the model's window, and must choose what to include. Zengram-lite makes this the
framework's job. Figure~\ref{fig:assemble} shows the whole path: the session-state
tables on the left, the six-phase assembly under a budget in the middle, and the
typed \code{ContextWindow} it returns on the right, with the measured
budget-eviction result of \S7.2 along the bottom.

\code{assembleContext(sessionId, branchId, budget, opts?)} runs a
\textbf{six-phase} assembly under a caller-supplied token budget. The phases, in
order, are system instructions, permission rules, knowledge, tasks, conversation
history (partitioned hot / warm / cold by recency), cross-branch summaries, and
pending state; \code{opts} sets each phase's share of the budget as a fraction. The
knowledge phase reads the session's project scope and the root scope, so facts
stored there are eligible for injection. History is packed newest-first until the
budget binds, and the remainder is summarized or skipped.

The return value is a structured \code{ContextWindow}, not a string: a list of
typed \textbf{blocks} (each with its type, content, token count, and the source
ids it came from), the total tokens used against the budget, counts of turns
\emph{included}, \emph{summarized}, and \emph{skipped}, and a
\textbf{\code{knowledgeFingerprint}} --- a hash that is stable for an unchanged
knowledge set. That fingerprint is a reuse signal: when it is unchanged between
turns, the knowledge portion of the prompt prefix is identical and a KV cache or a
cached prompt prefix can be reused. \S7.2 measures both behaviors --- the budget
binding and evicting oldest turns, and the fingerprint holding stable across
identical knowledge --- from the published artifact.

Storage-budget enforcement is the complementary operation:
\code{enforceBudget(scope)} prunes the oldest sessions when a scope exceeds its
configured storage budget, returning what it freed, so long-lived memory stays
bounded.

\subsection{Why this is the hard part}

Storing vectors is the easy half of agent memory; the value is in what surrounds
them. Turning a growing, branching conversation into a bounded, reusable prompt ---
under a schema that keeps the knowledge injected into that prompt consistent with
the facts the agent has confirmed, contradicted, and traced to their source --- is
the work an application otherwise does by hand and does incompletely. Making it a
framework operation, with a budget, a fingerprint, and typed provenance-bearing
blocks, is the session tier's contribution, and \S7.2 is its evidence.

\section{Running a Framework on a Synchronous Browser Surface}

A framework whose canonical operations call an LLM and an embedding model has a
problem in the browser that a pure database does not: those calls are
asynchronous, and the wasm surface it is invoked through is synchronous. This
section describes the mismatch and the seam that resolves it, plus the deployment
choices around it.

\subsection{The impedance mismatch}

Query execution in the engine is synchronous on the single wasm thread~[ZL~\S4.3],
so every method on the memory surface returns synchronously. But the framework's
most characteristic operations are inherently asynchronous in the browser:
computing an embedding means awaiting a model running in the tab (transformers.js,
ORT-Web) or a remote \code{/v1/embeddings} call, and extracting facts or
reflecting over memory means awaiting a completion model. A synchronous wasm
method cannot \code{await}. A naive port would either block the thread or require
an async wasm surface the engine does not have.

\subsection{The bring-your-own-result seam}

Zengram-lite resolves this by inverting the call. Rather than the framework
reaching out to an async model from inside a synchronous method, the application
computes the async result in JavaScript and hands the \emph{finished} result to
the framework, which then runs its normal synchronous code paths over it. The seam
has the same shape everywhere it appears:
\begin{itemize}[leftmargin=1.4em,itemsep=0.2em]
  \item \textbf{\code{rememberWithVector(subject, content, scope, vec)}} --- the
    application awaits its embedding model and passes the finished vector; the
    framework stores it with all the usual lifecycle and provenance handling.
    (\code{setEmbedFn} registers a \emph{synchronous} embedder for the simpler
    \code{remember}/\code{recall} path, which suits a lightweight or precomputed
    embedder; the \code{\ldots WithVector} pair is for real async models.)
  \item \textbf{\code{extractWithFacts(turnText, scope, sessionId, turnId,
    facts)}} --- the application calls its completion model to extract atomic facts
    from a turn and passes them in; the framework embeds each (via the registered
    embedder), dedupes and supersedes against existing knowledge, and stores them
    with session/turn provenance.
  \item \textbf{\code{reflectWithInsights(scope, insights, limit)}} --- the
    application's model synthesizes higher-level insights over the scope's
    knowledge and passes them in; the framework folds them into memory the same
    way.
\end{itemize}

The important property is that these are not three special cases but one design:
in every one, the async work is the application's and the framework work --- dedup,
supersession, embedding, provenance, storage --- is the framework's real code path,
unchanged from how it runs elsewhere. The framework is not stubbed in the browser;
it is fed. What v0.1 does \emph{not} do is call the model itself (\S8): extraction
and reflection are bring-your-own-result, not automatic, precisely because the
sync surface cannot make the async call.

\subsection{Engine modes: isolated or shared}

Memory needs an engine, and the application chooses whether it gets its own or
shares one. \code{ZengramMemory.open(dim)} opens memory over a fresh, isolated
engine with a private catalog --- the right choice when the page needs only
memory, or wants memory's storage in a separate trust domain from any SQL it runs.
\code{ZengramMemory.overEngine(db, dim)} builds memory over an existing
\code{ZetaDb}'s engine --- one wasm heap, one catalog --- so the application's SQL
tables and memory's tables live in one database and can be queried together. Both
are first-class. In shared mode memory reserves the table names its migrations
create, so the application avoids colliding with them; the migrations are
idempotent (tracked in \code{\_zengram\_migrations}), so attaching memory over an
already-initialized or snapshot-restored engine is safe and cheap. The reserved set
is 46 names in the current build; \S\ref{sec:schema} groups them and points to the
authoritative list rather than reproducing it.

\subsection{Persistence}

Durability is the engine's snapshot model~[ZL~\S5], applied to the memory database:
\code{exportSnapshot()} serializes the whole database --- in shared mode including
the application's SQL tables, since they are one database --- to a
\code{Uint8Array}, and \code{ZengramMemory.openFromSnapshot(bytes, dim)} rehydrates
it. There is no incremental save; each snapshot is $O(\text{database size})$. The
shipped browser agent (\code{demo/agent/}) demonstrates the production pattern: a
dirty flag set only by memory-writing turns, a $\sim$60-second throttle so a run of
new facts coalesces into a single save, and a flush on tab hide or close. A clean
quit loses nothing; a hard crash loses at most one throttle window of facts.
Snapshots are origin-agnostic bytes --- the application stores them in OPFS,
IndexedDB, or ships them elsewhere.

\section{Evaluation: Coverage, Framework Behavior, and Size}

We evaluate the framework, not the engine. The engine's throughput, concurrency,
and soak behavior are zeta-lite's~[ZL] and are inherited unchanged; re-running them
here would measure the same engine. What this section establishes is that the
framework layer \emph{works, completely, in the target environment} --- shown by
harnesses that drive the real artifact --- that its distinctive session-tier
behavior is as described (\S7.2), and that the whole superset fits the size
envelope claimed (\S7.3). All measurements run against the \textbf{published}
artifact and reproduce from the public repository.

\subsection{Functional coverage}

Four harnesses drive the compiled artifact through the framework's surface:
\begin{itemize}[leftmargin=1.4em,itemsep=0.2em]
  \item \textbf{\code{smoke.mjs}} exercises the memory core and the re-exported
    engine surface --- remember and recall (with a toy embedder),
    confirm/contradict moving confidence, decay, snapshot export-and-restore
    preserving recall, peer-attributed facts, and raw \code{query()} over the
    memory tables. \textbf{19 checks, all passing.}
  \item \textbf{\code{smoke\_agent.mjs}} walks the full agent loop against the same
    artifact --- \code{createSession}, \code{appendTurn}/\code{addPart},
    \code{recordToolCall}/\code{completeToolCall}, \code{assembleContext}, and the
    \code{extractWithFacts} / \code{reflectWithInsights} bridges. \textbf{27
    checks, all passing.}
  \item \textbf{\code{smoke\_model.mjs}} replaces the toy embedder with real
    all-MiniLM-L6-v2 vectors and confirms \textbf{zero-keyword-overlap recall}:
    queries sharing no words with the stored facts retrieve the right ones, which
    is the property that distinguishes semantic recall from text matching.
  \item \textbf{\code{e2e\_ops.mjs}} drives the real demo page in headless Chromium
    --- the DOM path the Node smokes cannot reach --- typing facts into the form,
    clicking the confirm/contradict and decay controls, and asserting the
    confidence bars move.
\end{itemize}

Together these cover every tier: the knowledge lifecycle, the
session/turn/tool-call model, context assembly, the bring-your-own-result bridges,
persistence, and the real embedding path, each against the artifact the demos
actually load.

\subsection{Context assembly under a token budget}

This is the session tier's distinctive claim, and it is directly reproducible. We
seed a session with five knowledge facts and twelve turns, then call
\code{assembleContext} at three budgets and observe what it packs:

\begin{table}[htbp]
  \centering
  \begin{tabular}{r r r r l}
    \toprule
    \textbf{Token budget} & \textbf{Total tokens} & \textbf{Turns} & \textbf{Turns} & \textbf{Blocks} \\
                          & \textbf{packed}       & \textbf{included} & \textbf{skipped} & \\
    \midrule
    200  & 141 & 3  & 9 & system, knowledge, 3$\times$ turn \\
    1000 & 579 & 12 & 0 & system, knowledge, 12$\times$ turn \\
    4000 & 579 & 12 & 0 & system, knowledge, 12$\times$ turn \\
    \bottomrule
  \end{tabular}
\end{table}

At a tight budget of 200 tokens the assembly includes the system and knowledge
blocks and only the three most recent turns, \textbf{skipping the other nine} ---
the budget is the binding constraint and it evicts oldest history first. Given room
(budget 1000), all twelve turns fit at 579 tokens; raising the budget to 4000
changes nothing, since the material, not the budget, is now the limit. Across
repeated calls on the \textbf{unchanged} knowledge set, the
\code{knowledgeFingerprint} is byte-for-byte identical, confirming the reuse signal
is stable: an agent can compare fingerprints between turns and reuse a cached prompt
prefix when knowledge has not changed. Both behaviors --- budget-bound eviction and
fingerprint stability --- are what \S\ref{sec:assemble} describes, measured on the
published bundle.

\subsection{Artifact size}

The published \code{zengram\_wasm\_bg.wasm} is \textbf{10,400,328 bytes raw /
2.95\,MB gzipped}. The comparison that matters is against the engine it supersets:
zeta-lite is 2.87\,MB gzipped~[ZL~\S7.4]. The entire agentic-memory framework ---
the knowledge lifecycle, the session/turn/tool-call model, context assembly, and
the bring-your-own-result bridges --- therefore adds on the order of
\textbf{0.08\,MB gzip} on top of the SQL engine already present. This is the
concrete meaning of the superset design: a page that would load zeta-lite for its
database can load zengram-lite instead and gain the whole memory framework for a
few percent more transfer, with no second engine and no second download.

\subsection{End to end: the shipped local agent}

The strongest evidence that the framework is a working system, not a set of passing
unit tests, is the local agent shipped in \code{demo/agent/}. Its loop, its tools
(memory, OPFS files, web fetch), and its memory all run in the browser tab; only
LLM inference is a remote call to an OpenAI-compatible endpoint the user
configures. It stores facts as it learns them, recalls them by meaning on later
turns, records its tool calls, and snapshots its memory to OPFS on a throttle ---
so the memory survives a full browser restart, not merely a reload. Everything \S4
through \S6 describes runs in that page, driven by a real model, which is the
framework operating as intended.

\section{Limitations and Future Work}

Zengram-lite is a v0.1 preview, and its boundaries are deliberate and stated.

\textbf{Extraction and reflection are bring-your-own-result, not automatic.}
Because the wasm surface is synchronous and the completion model is asynchronous
(\S6.1), the framework cannot call the model itself; the application supplies the
extracted facts and synthesized insights. Correspondingly, \code{remember} stores
content verbatim and does not split prose into atomic facts on its own. When an
async bridge lands, automatic extraction becomes possible; in v0.1 it is the
caller's step.

\textbf{Much of the compiled framework is not yet exposed as a typed surface.} The
build carries more than the JS API currently exposes: episodic timelines
(\code{session\_timeline}, \code{turns\_since}), provenance tracing
(\code{traceBack}/\code{traceForward}), the event log, reminders, permission rules,
turn summarization, and in-browser memory branching
(\code{fork}/\code{merge}/\code{rebase}). These are reachable today only by
querying their tables with \code{query()}; a wider typed JS surface is planned.
(Fleet coordination, pgwire, and object-knowledge ingest are excluded from the
browser build by design --- they belong server-side and pull native dependencies.)

\textbf{Durability is snapshot-based, and inherited.} A committed change is durable
only as of the last snapshot the application persisted; there is no incremental
save, and each snapshot is $O(\text{database size})$. This is zeta-lite's
persistence model~[ZL~\S5.3], with its window; the related engine limitation that
snapshots do not yet capture branches~[ZL~\S8] applies to memory branching too.

\textbf{Framework evidence is functional, not throughput.} This paper measures
coverage, context-assembly behavior, and size; it adds no framework
micro-benchmarks. Engine performance --- where the operations' cost actually lives
--- is zeta-lite's~[ZL], and the memory tiers are ordinary tables and queries over
that engine.

\textbf{Future work} centers on exposing the compiled-in surface (episodic
timelines, provenance tracing, memory branching) as typed APIs, wiring automatic
extraction when an async model bridge is available, and --- beyond the teaser ---
the full Zengram framework this build fronts (planned for open-source release under
Apache-2.0) and the closed Zeta engine beneath it.

\section{Conclusion}

The client side has had vector stores for a while; what it has lacked is a memory
\emph{framework} --- the lifecycle, the session model, and the context assembly
that make stored vectors behave like an agent's memory rather than a search index.
Zengram-lite is that framework, compiled to run entirely in a browser tab. A fact
in it has confidence that moves and importance that decays; the conversation is
modeled as sessions, turns, typed parts, and tool calls rather than an opaque
array; and each turn's prompt is assembled under a token budget with a fingerprint
that says when it can be reused. All of it sits on the zeta-lite engine~[ZL], from
which it inherits transactional consistency across every tier and branchable memory
for the cost of a timestamp, and all of it ships in one $\sim$2.95\,MB artifact ---
the whole framework for a few percent more than the SQL engine alone. The one
genuinely new piece of engineering the browser forced, the bring-your-own-result
seam, is what lets a framework built around an LLM run on a surface that cannot call
one. Zengram-lite is the smallest and most complete public form of the full
Zengram framework, and the in-browser teaser for it; the engine beneath it stands
on its own in the companion report.

\section*{References}
\addcontentsline{toc}{section}{References}

{\small
\emph{The items marked as zengram-lite's own artifacts are reproducible from the
public repository.}

\begin{description}[leftmargin=3.2em,style=nextline,itemsep=0.3em]
  \item[{[ZL]}] G. Zhang. \emph{Zeta-Lite: A Concurrent, Branchable In-Browser SQL
    Database for Agentic Memory.} arXiv:2609.01818, 2026.
    \url{https://arxiv.org/abs/2609.01818}. Repository:
    \url{https://github.com/genezhang/zeta-lite}. The account
    of the engine underneath zengram-lite --- MVCC, overlapping snapshot isolation,
    copy-on-write branching, snapshot-to-OPFS persistence, the wasm-not-WASI host
    binding, and the throughput/concurrency/soak evaluation.
  \item[{[1]}] C. Packer, S. Wooders, K. Lin, V. Fang, S. G. Patil, I. Stoica, and
    J. E. Gonzalez. \emph{MemGPT: Towards LLMs as Operating Systems.}
    arXiv:2310.08560, 2023. \url{https://arxiv.org/abs/2310.08560}. See also Letta
    (letta.com), the successor framework.
  \item[{[2]}] P. Chhikara, D. Khant, S. Aryan, T. Singh, and D. Yadav. \emph{Mem0:
    Building Production-Ready AI Agents with Scalable Long-Term Memory.}
    arXiv:2504.19413, 2025. \url{https://arxiv.org/abs/2504.19413}. Implementation:
    \url{https://github.com/mem0ai/mem0}.
  \item[{[3]}] LangChain (python.langchain.com) and LlamaIndex (llamaindex.ai) ---
    conversational-memory and retrieval/context-construction frameworks for LLM
    applications (accessed 2026-08-31).
  \item[{[4]}] zengram-lite. \emph{API reference and behavioral docs}
    (\code{docs/api.md}, \code{docs/how-it-works.md}, \code{docs/engine-modes.md})
    and \emph{smoke / e2e harnesses} (\code{demo/smoke.mjs} --- 19 checks;
    \code{demo/smoke\_agent.mjs} --- 27 checks; \code{demo/smoke\_model.mjs} ---
    real all-MiniLM recall; \code{demo/e2e\_ops.mjs} --- headless-Chromium DOM).
    All run against the published artifact.
    \url{https://github.com/genezhang/zengram-lite}.
\end{description}
}

\end{document}